# High-order integral-chain differentiator and application to acceleration feedback

Xinhua Wang

Advanced Control Technology, Department of Electrical & Computer Engineering, National University of Singapore, Singapore 117576

E-mail: wangxinhua04@gmail.com

**Abstract:** The equivalence between integral-chain differentiator and usual high-gain differentiator is given under suitable coordinate transformation. Integral-chain differentiator can restrain noises more thoroughly than usual high-gain linear differentiator. In integral-chain differentiator, disturbances only exist in the last differential equation and can be restrained through each layer of integrator. Moreover, a nonlinear integral-chain differentiator is designed which is the expansion of linear integral-chain differentiator. Finally, a 3-order differentiator is applied to the estimation of acceleration for a second-order uncertain system.

**Keywords:** Integral-chain differentiator, high-gain differentiator, acceleration.

## I. Introduction

Obtaining the velocities and accelerations of tracked targets is crucial for several kinds of systems with correct and timely performances, such as the missile-interception systems [1] and underwater vehicle systems [2], and in which disturbances must be restrained. The accelerations were seldom used to feedback control because they are difficult to be obtained. Therefore, the robustness of differentiators should be taken into consideration.

Differentiation of signals is an old and well-known problem [3]–[5] and has attracted more attention in recent years [6]–[9]. In [6, 7], a differentiator via second-order (or high-order) sliding modes algorithm has been proposed. The information one needs to know on the signal is the upper bounds for Lipschitz constant of the derivatives of the signal. It constrains the types of input signals. And for this differentiator, the chattering phenomenon is inevitable. The popular high-gain differentiators in [8, 9] provide for an exact derivative when their gains tend to infinity. Noises exist in each layer of differential equations, therefore, all the derivatives estimations of input signal are all affected by noises directly.

In [10], a finite-time-convergent differentiator based on singular perturbation technique has been presented. However, the differentiators are complicated and difficult to be implemented in practice, due to the long computation time. In [11], a nonlinear tracking-differentiator with high speed is designed, and it succeeds in application to velocity estimation for low-speed regions only based on position measurements [12], in which only the convergence of signal tracking is described for this differentiator, but the convergence of derivative tracking is not given.

In this paper, we give the equivalence between integral-chain differentiator and use high-gain differentiator under suitable coordinate transformation. Because the structure of high-order integral chains exists in differentiator, and noises only exist in the last differential equation, each layer integrator can restrain noises. Therefore, integral-chain differentiator can restrain noises more thoroughly than usual high-gain linear differentiator. A nonlinear integral-chain differentiator is designed which is the expansion of linear integral-chain differentiator to increase convergent velocity. Moreover, a 3-order differentiator is applied to the estimation of acceleration for a second-order uncertain system.

## II. Preliminaries

First of all, the concepts related to finite-time stability are given [13]-[17].

***Definition 1 [13]:*** Consider a time-invariant system in the form of

$$\dot{x} = f(x) \qquad f(0) = 0, \quad x \in \Re^n \tag{1}$$

where $f : D_0 \to \Re^n$ is continuous on open neighborhood $D_0 \subseteq \Re^n$ of the origin. The equilibrium $x=0$ of the system is (locally) finite-time stable if (i) it is asymptotically stable, in $D$, an open neighborhood of the origin, with $D \subseteq D_0$; (ii) it is finite-time convergent in $D$, that is, for any initial condition $x_0 \in D \setminus \{0\}$, there is a settling time $T > 0$ such that every solution $x_0(t,t_0)$ of system (1) is defined with $x_0(t,t_0) \in D \setminus \{0\}$ for $t \in [0,T)$ and satisfies $\lim_{t \to T} x(t,t_0) = 0$ and $x(t,t_0) = 0$, if $t \geq T$. Moreover, if $D = \Re^n$, the origin $x = 0$ is globally finite-time stable.

***Definition 2 [14]-[16]:*** A family of dilations of $\delta_\rho^r$ is a mapping that assigns to every real $\rho > 0$ a diffeomorphism

$$\delta_\rho^r(x_1,\cdots,x_n) = (\rho^{r_1} x_1,\cdots,\rho^{r_n} x_n), \qquad (2)$$

where $x_1,\cdots,x_n$ are suitable coordinates on $\Re^n$ and $r = (r_1,\cdots,r_n)$ with the dilation coefficients $r_1,\cdots,r_n$ positive real numbers. A vector $f(x) = (f_1(x),\cdots,f_1(x))^T$ is homogeneous of degree $k \in \Re$ with respect to the family of dilations $\delta_\rho^r$ if

$$f_i(\rho^{r_1} x_1,\cdots,\rho^{r_n} x_n)^T = \rho^{k+r_i} f(x), \quad i=1,\cdots,n, \quad \rho > 0 \qquad (3)$$

***Theorem [17]:*** The origin is a finite-time stable equilibrium of $f$ if and only if the origin is an asymptotically stable equilibrium of $f$ and $k < 0$.

***Denotations:***

$sig(y)^\alpha = |y|^\alpha \operatorname{sgn}(y)$, $\alpha > 0$. It is obvious that $sig(y)^\alpha = y^\alpha$ only if $\alpha = q/p$. where $p,q$ are positive odd numbers. Moreover,

$$\frac{d}{dy}|y|^{\alpha+1} = (\alpha+1)sig(y)^\alpha, \frac{d}{dy}sig(y)^{\alpha+1} = (\alpha+1)|y|^\alpha \qquad (4)$$

### III. Equivalence of integral-chain linear differentiator and usual high-gain linear differentiator

We know that linear high-gain differentiator is [8, 9]:

$$\begin{aligned}
\dot{w}_1 &= w_2 - \frac{a_n}{\varepsilon}(w_1 - v(t)) \\
\dot{w}_2 &= w_3 - \frac{a_{n-1}}{\varepsilon^2}(w_1 - v(t)) \\
&\cdots \\
\dot{w}_{n-1} &= w_n - \frac{a_2}{\varepsilon^{n-1}}(w_1 - v(t)) \\
\dot{w}_n &= -\frac{a_1}{\varepsilon^n}(w_1 - v(t))
\end{aligned} \qquad (5)$$

where $(w_1,\cdots,w_n)$ is the state of high-gain differentiator, and $a_i, i=1,\cdots,n$, are given such that $s^n + a_n s^{n-1} + \cdots + a_2 s + a_1 = 0$ is Hurwitz. For high-gain linear differentiator (5), we can have that

$$\lim_{\varepsilon \to 0} w_i = v^{(i-1)}(t), \quad i = 1, \cdots, n \tag{6}$$

An integral-chain differentiator is designed as:

$$\begin{aligned}
\dot{x}_1 &= x_2 \\
\dot{x}_2 &= x_3 \\
&\cdots \\
\dot{x}_{n-1} &= x_n \\
\dot{x}_n &= -\frac{a_1}{\varepsilon^n}(x_1 - v(t)) - \frac{a_2}{\varepsilon^{n-1}} x_2 - \cdots - \frac{a_{n-1}}{\varepsilon^2} x_{n-1} - \frac{a_n}{\varepsilon} x_n
\end{aligned} \tag{7}$$

where the parameters are the same with those in high-gain differentiator (5). We can also obtain the results of

$$\lim_{\varepsilon \to 0} x_i = v^{(i-1)}(t), \quad i = 1, \cdots, n \tag{8}$$

In the following, we will give the equivalence between (5) and (7).

***Theorem 1:*** High-gain differentiator (5) and integral-chain differentiator are equivalent if the following relations are satisfied.

$$a_{i+1}^2 = a_i a_{i+2}, \quad i = 1, \cdots, n-2 \tag{9}$$

***Proof:*** Let

$$\begin{aligned}
x_1 &= w_1 - \varepsilon \frac{a_2}{a_1} w_2 \\
x_2 &= w_2 - \varepsilon \frac{a_3}{a_2} w_3 \\
&\cdots \\
x_{n-1} &= w_{n-1} - \varepsilon \frac{a_n}{a_{n-1}} w_n \\
x_n &= w_n
\end{aligned} \tag{10}$$

then from (7) and (10) we can obtain that

$$\begin{aligned}
\dot{x}_n = \dot{w}_n = &-\frac{a_1}{\varepsilon^n}\left(w_1 - \varepsilon \frac{a_2}{a_1} w_2 - v(t)\right) - \frac{a_2}{\varepsilon^{n-1}}\left(w_2 - \varepsilon \frac{a_3}{a_2} w_3\right) \\
&- \cdots - \frac{a_{n-1}}{\varepsilon^2}\left(w_{n-1} - \varepsilon \frac{a_n}{a_{n-1}} w_n\right) - \frac{a_n}{\varepsilon} w_n
\end{aligned} \tag{11}$$

Therefore, we have

$$\dot{w}_n = -\frac{a_1}{\varepsilon^n}(w_1 - v(t)) \tag{12}$$

From (10) and (7), we have $\dot{x}_{n-1} = \dot{w}_{n-1} - \varepsilon \frac{a_n}{a_{n-1}} \dot{w}_n = w_n$, then we get

$$\dot{w}_{n-1} = w_n - \frac{1}{\varepsilon^{n-1}}\left(\frac{a_1 a_n}{a_{n-1}}\right)(w_1 - v(t)) \tag{13}$$

From (9), we can get

$$\frac{a_1 a_n}{a_{n-1}} = \frac{a_1 a_{n-1} a_n}{a_{n-1}^2} = \frac{a_1 a_{n-1} a_n}{a_{n-2} a_n} = \frac{a_1 a_{n-1}}{a_{n-2}} = \cdots = \frac{a_1 a_2}{a_1} = a_2 \tag{14}$$

Therefore, (13) can be written as

$$\dot{w}_{n-1} = w_n - \frac{a_2}{\varepsilon^{n-1}}\left(w_1 - v(t)\right) \tag{15}$$

From (10) and (7), we have $\dot{x}_{n-2} = \dot{w}_{n-2} - \varepsilon \frac{a_{n-1}}{a_{n-2}} \dot{w}_{n-1} = x_{n-1}$, then we get

$$\dot{w}_{n-2} = w_{n-1} - \varepsilon \frac{a_n}{a_{n-1}} w_n + \varepsilon \frac{a_{n-1}}{a_{n-2}}\left(w_n - \frac{1}{\varepsilon^{n-1}}\left(\frac{a_1 a_n}{a_{n-1}}\right)\left(w_1 - v(t)\right)\right) \tag{16}$$

Because of $a_{n-1}^2 = a_{n-2} a_n$ in (9), we have

$$\dot{w}_{n-2} = w_{n-1} - \frac{1}{\varepsilon^{n-2}}\left(\frac{a_1 a_n}{a_{n-2}}\right)\left(w_1 - v(t)\right) \tag{17}$$

From (9), we can get

$$\frac{a_1 a_n}{a_{n-2}} = \frac{a_1 a_{n-2} a_n}{a_{n-2}^2} = \frac{a_1 a_{n-2} a_n}{a_{n-3} a_{n-1}} = \frac{a_1 a_{n-1}}{a_{n-3}} = \cdots = a_3 \tag{18}$$

Therefore, we have

$$\dot{w}_{n-2} = w_{n-1} - \frac{a_3}{\varepsilon^{n-2}}\left(w_1 - v(t)\right) \tag{19}$$

Finally, from (10) and (7), we have $\dot{x}_1 = \dot{w}_1 - \varepsilon \frac{a_2}{a_1} \dot{w}_2 = x_2$, then we can get

$$\dot{w}_1 = w_2 - \varepsilon \frac{a_3}{a_2} w_3 + \varepsilon \frac{a_2}{a_1}\left(w_3 - \frac{1}{\varepsilon^2}\left(\frac{a_1 a_n}{a_2}\right)\left(w_1 - v(t)\right)\right) \tag{20}$$

Because of $a_2^2 = a_1 a_3$ in (9), we have

$$\dot{w}_1 = w_2 - \frac{a_n}{\varepsilon}\left(w_1 - v(t)\right) \tag{21}$$

Then we get high-gain differentiator (5). This concludes the proof. □

From Theorem 1, integral-chain linear differentiator has the same tracking results with that of usual high-gain differentiator in the condition that $\varepsilon$ is sufficiently small.

Usually there are noises in the input signal. From (7), noises only exist in the last layer of differential equation. Therefore, noises are retrained by layers of integral chains thoroughly. However, noises exist in every layers of usual high-gain differentiator. The noise cannot be restrained thoroughly. Therefore, integral-chain differentiator has better ability of restraining noises than usual high-gain differentiator.

### IV. Nonlinear integral-chain differentiator

We first give nonlinear integral-chain differentiator as follow:

$$\begin{aligned}
\dot{x}_1 &= x_2 \\
\dot{x}_2 &= x_3 \\
&\cdots \\
\dot{x}_{n-1} &= x_n \\
\dot{x}_n &= -\frac{1}{\varepsilon^n}\left[a_1 sig(x_1 - v(t))^{\alpha_1} + \sum_{i=2}^{n} a_i sig(\varepsilon^{i-1} x_2)^{\alpha_i}\right]
\end{aligned} \tag{22}$$

A theorem about nonlinear integral-chain differentiator is given in the following.

***Theorem 2:*** For nonlinear integral-chain differentiator, we have that: there exist $\gamma > 0$ (where $\rho\gamma > 0$) and $\Gamma > 0$ such that

$$x_i - v^{(i-1)}(t) = O(\varepsilon^{\rho\gamma-i+1}), \quad i = 1, \cdots, n \tag{23}$$

for $t \geq \varepsilon\Gamma$. Where $\varepsilon > 0$ is the perturbation parameter and $O(\varepsilon^{\rho\gamma-i+1})$ denotes the approximation of $\varepsilon^{\rho\gamma-i+1}$ order [18] between $x_i$ and $v^{(i-1)}(t)$; and $\gamma = (1-\theta)/\theta \in (0, \min\{\rho/\rho+n, 1/2\})$. $a_i, i = 1, \cdots, n$, are given such that $s^n + a_n s^{n-1} + \cdots + a_2 s + a_1 = 0$ is Hurwitz.

In order to prove Theorem 2, we will give a lemma in the following.

***Lemma 1:*** The equilibrium $z = 0$ of system

$$\begin{aligned}
\dot{z}_1 &= z_2 \\
\dot{z}_2 &= z_3 \\
&\cdots \\
\dot{z}_2 &= z_3 \\
\dot{z}_3 &= -\sum_{i=1}^{n} a_i sig(z_i)^{\alpha_i}
\end{aligned} \tag{24}$$

is finite-time stable. Where $s^n + a_n s^{n-1} + \cdots + a_2 s + a_1 = 0$ is Hurwitz, $0 < \alpha_1 < 1$, $\alpha_i = \dfrac{n\alpha_1}{(i-1)\alpha_1 + (n-i+1)}$, $i = 2, \cdots, n$.

***Proof:*** From [19], we know that system (24) is asymptotically stable. In the following, we will prove that the equilibrium $z = 0$ of system (24) is finite-time stable. From definition 2 and (24), we have

$$\rho^{r_i} z_i = \rho^{k+r_{i-1}} z_i, \quad i = 2, \cdots, n \tag{25}$$

and

$$-\sum_{i=2}^{n} a_{i-1} sig(\rho^{r_{i-1}} z_{i-1})^{\alpha_{i-1}} = \rho^{k+r_n}\left[-\sum_{i=2}^{n} a_{i-1} sig(z_{i-1})^{\alpha_{i-1}}\right] \tag{26}$$

Therefore, (25) and (26) can be written respectively as:

$$r_i = k + r_{i-1}, \quad i = 2, \cdots, n \tag{27}$$

$$r_{i-1}\alpha_{i-1} = k + r_n, \quad i = 2, \cdots, n \tag{28}$$

Therefore, from (27), we have

$$r_i = (i-1)k + r_1, \quad i = 2, \cdots, n \tag{29}$$

and from (29), relation (29) can be written as:

$$r_1\alpha_1 = \left[(i-1)k + r_1\right]\alpha_i = nk + r_1, \quad i = 2, \cdots, n \tag{30}$$

Therefore, from $r_1\alpha_1 = \left[(i-1)k + r_1\right]\alpha_i$ in (30), we have

$$r_1 = \frac{(i-1)k\alpha_i}{\alpha_1 - \alpha_i}, \quad i = 2, \cdots, n \tag{31}$$

And from $r_1\alpha_1 = nk + r_1$ in (30), we get

$$r_1 = \frac{nk}{\alpha_1 - 1} \tag{32}$$

Therefore, from (31) and (32), we have

$$\frac{(i-1)k\alpha_i}{\alpha_1 - \alpha_i} = \frac{nk}{\alpha_1 - 1}, \quad i = 2, \cdots, n \tag{33}$$

From (33), we get

$$\alpha_i = \frac{n\alpha_1}{(i-1)\alpha_1 + (n-i+1)}, \quad i = 2, \cdots, n \tag{34}$$

Due to $0 < \alpha_1 < 1$, and from (34), we have

$$0 < \alpha_{i-1} < \alpha_i < 1, \quad i = 2, \cdots, n \tag{35}$$

Moreover, we can get

$$r_i = (i-1)k + r_1 = (i-1)k + \frac{nk}{\alpha_1 - 1} = \frac{\left[(\alpha_1 - 1)(i-1) + n\right]k}{\alpha_1 - 1}, \quad i = 2, \cdots, n \tag{36}$$

Because $\alpha_1 \in (0,1)$, $r_{i-1} > 0, i = 2, \cdots, n$, and from (36), we have

$$k < 0 \tag{37}$$

Therefore, from definition 2 and Theorem 2 in [17], the equilibrium $z = 0$ of system (24) is finite-time stable. □

From Theorem 1 in [10], we can get that (23) is satisfied for nonlinear integral-chain differentiator.

**Remark 1:** For nonlinear integral-chain differentiator (22), if we let $\alpha_i = 1, i = 1, \cdots, n$, then we can obtain linear integral-chain differentiator (7). And we can combine (2) and (22) to form a hybrid integral-chain differentiator as follow:

$$\dot{x}_1 = x_2$$
$$\ldots$$
$$\dot{x}_{n-1} = x_n$$
$$\dot{x}_n = -\frac{1}{\varepsilon^n}\left[a_1(x_1 - v(t)) + \sum_{i=2}^{n} a_i \varepsilon^{i-1} x_2 + b_1 sig(x_1 - v(t))^{\alpha_1} + \sum_{i=2}^{n} b_i sig(\varepsilon^{i-1} x_2)^{\alpha_i}\right] \tag{31}$$

The nonlinear items are introduced into the linear integral-chain differentiator. The convergences of linear, nonlinear and hybrid integral-chain differentiator are shown in Fig. 1, Fig. 2 and Fig. 3. From Fig. 3, we can see that high-speed convergence is guaranteed in the whole course for hybrid integral-chain differentiator.

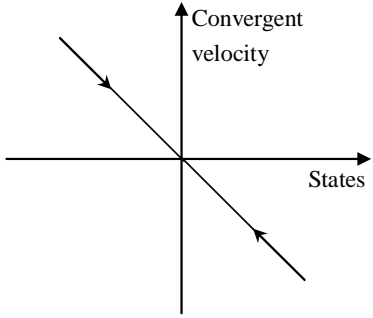
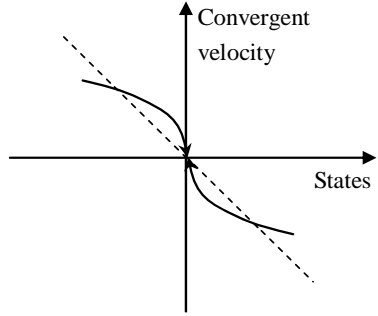
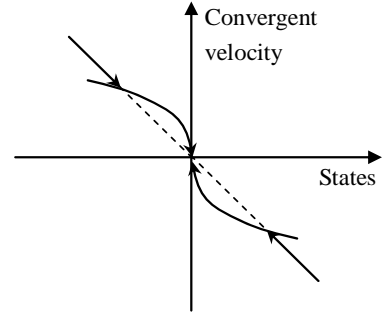

Fig.1. Convergent velocity in linear form   Fig.2. Convergent velocity in nonlinear form   Fig. 3. Convergent velocity in hybrid form

## V. Control based on 3-order differentiator for uncertain system with output noises

The following system is considered:
$$\dot{\theta} = \omega$$
$$\dot{\omega} = f(t) + bu \tag{32}$$
$$y_{ouput} = \theta + \delta$$

where $f(t)$ is the uncertainty, $b \in \Re$ is constant, measurement output $\theta$ with noise $\delta$, the reference trajectory $\theta_d$. Let $e_1 = \theta - \theta_d, e_2 = \omega - \dot{\theta}_d$. Therefore, the tracking error system is

$$\dot{e}_1 = e_2$$
$$\dot{e}_2 = f(t) - \ddot{\theta}_d + bu \tag{33}$$

The sliding variable is selected as $s = e_2 + k_u e_1$. The Lyapunov function is selected as

$$V = \frac{1}{2}s^2 \tag{34}$$

Therefore, we have

$$\dot{V} = s\left(f(t) - \ddot{\theta}_d + bu + k_u e_2\right) \tag{35}$$

If the up boundness of $f(t)$ is known, we can select the control input $u$ as

$$u = \frac{1}{b}\left[-k_u e_2 + \ddot{\theta}_d - l\,\text{sgn}(s)\right] \tag{36}$$

where $k_u > 0$, and $|f(t)| < l$.

If $\omega$ and the up boundness of $f(t)$ are unknown, we need to estimate $f(t)$ and $\omega$. In the following, we adopt respectively 3-order integral-chain and usual high-gain differentiators to estimate $f(t)$ and $\omega$ and use them to feedback control. 3-order integral-chain and usual high-gain differentiators are shown respectively in the following:

$$\begin{aligned}\dot{x}_1 &= x_2 \\ \dot{x}_2 &= x_3 \\ \dot{x}_3 &= -\frac{a_1}{\varepsilon^3}(x_1 - \theta) - \frac{b_1}{\varepsilon^2}x_2 - \frac{c_1}{\varepsilon}x_3\end{aligned} \quad (37)$$

and

$$\begin{aligned}\dot{w}_1 &= w_2 - \frac{a_3}{\varepsilon}(w_1 - \theta) \\ \dot{w}_2 &= w_3 - \frac{a_2}{\varepsilon^2}(w_1 - \theta) \\ \dot{w}_3 &= -\frac{a_1}{\varepsilon^3}(w_1 - \theta)\end{aligned} \quad (38)$$

From the measurement value $\theta$ with noise, we use the differentiator to estimate the first-order and second-order derivatives of $\theta$, i.e., $\hat{\omega}$ and $\hat{\dot{\omega}}$. Therefore, we can obtain

$$\hat{f}(t) = \hat{\dot{\omega}} - bu \quad (39)$$

In $\hat{f}(t)$, there is the information of acceleration. Accordingly, the control input $u$ based on 3-order differentiator can be designed as

$$u = \frac{1}{b}\left[-k_u(\hat{\omega} - \dot{\theta}_d) + \ddot{\theta}_d - l\,\mathrm{sgn}(s) - \hat{f}(t)\right] \quad (40)$$

where $|\delta|<l$, $s = (\hat{\omega} - \dot{\theta}_d) + k_u(\hat{\theta} - \theta_d)$.

## VI. Simulation

In this section, we use a numerical simulation to illustrate the effectiveness of the proposed method for the uncertain system (32). The goal is of forcing the output $\theta$ to track a sinusoid signal of $\theta_d=\sin t$. Parameters: $b=133$, $f(t)=-25x_2$, $k_u=10$, $l=0.15$. $a_1=10$, $b_1=10$, $c_1=10$, $\varepsilon=0.01$, $|\delta| \leq 0.05$.

The tracking curves comparison between integral-chain and usual high-gain differentiators are shown in Fig.4 and Fig. 5. It is seen that the tracking errors asymptotically converge to the desired trajectory, and the controller $u(t)$ is bounded. From the comparison, we can find that the better estimating and tracking effects are obtained by integral-chain differentiator. More noises can be restrained by integral-chain differentiator.

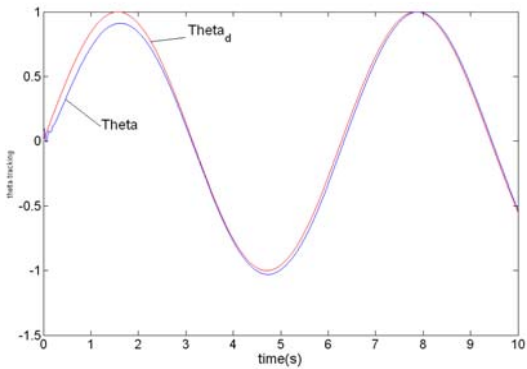

Fig.4-1 $\theta$ tracking $\theta_d$ with integral-chain differentiator

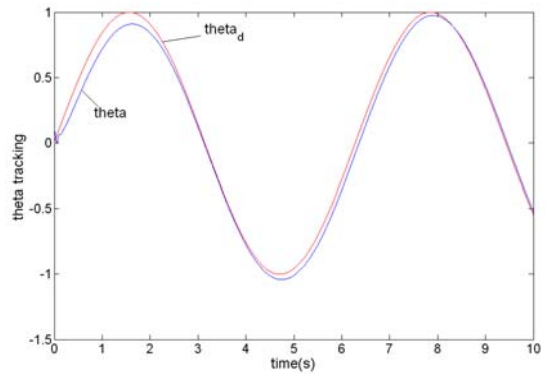

Fig.5-1 $\theta$ tracking $\theta_d$ with usual high-gain differentiators

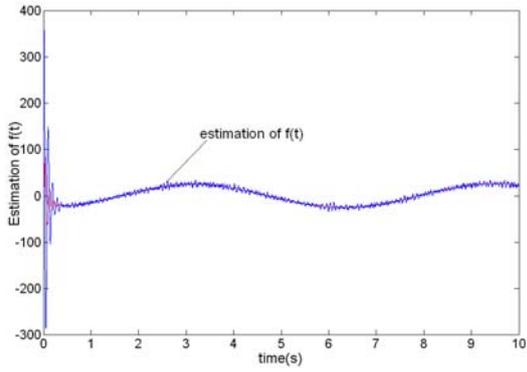

Fig. 4-2 Estimation of *f*(*t*) with integral-chain differentiator

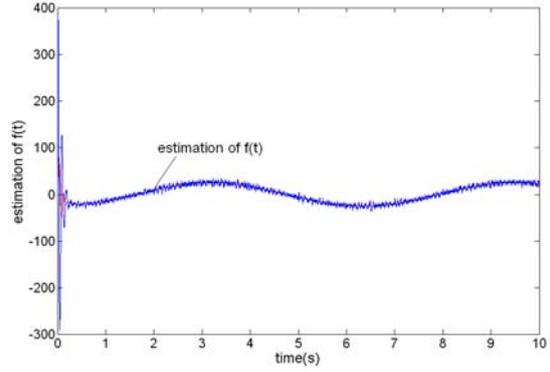

Fig. 5-2 Estimation of *f*(*t*) with usual high-gain differentiators

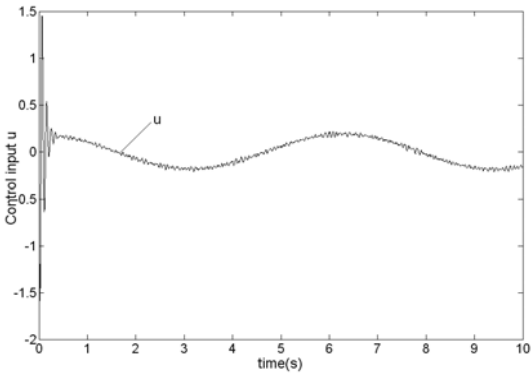

Fig. 4-3 Control input *u* with integral-chain differentiator

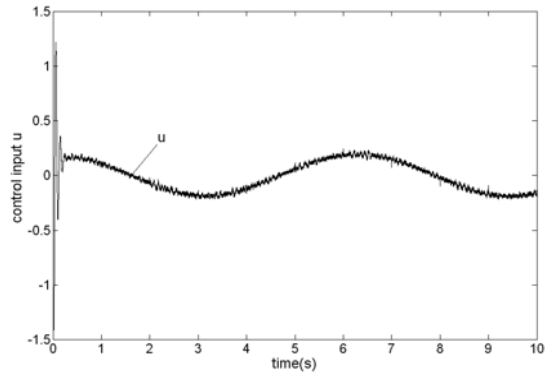

Fig. 5-3 Control input *u* with usual high-gain differentiators

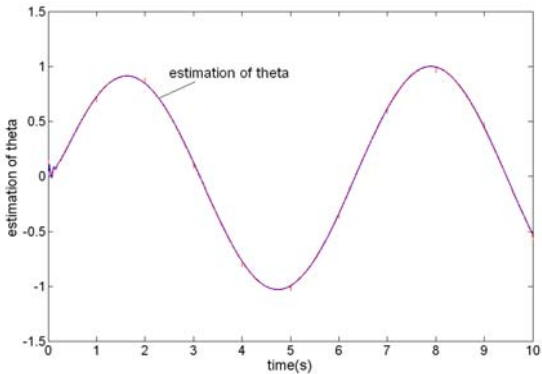

Fig. 4-4 $\hat{\theta}$ tracking $\theta_d$ by integral-chain differentiator

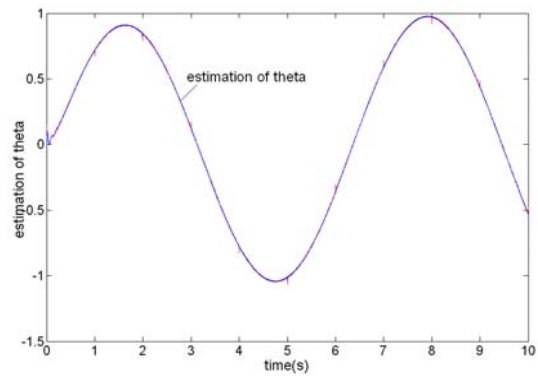

Fig. 5-4 $\hat{\theta}$ tracking $\theta_d$ by usual high-gain differentiators

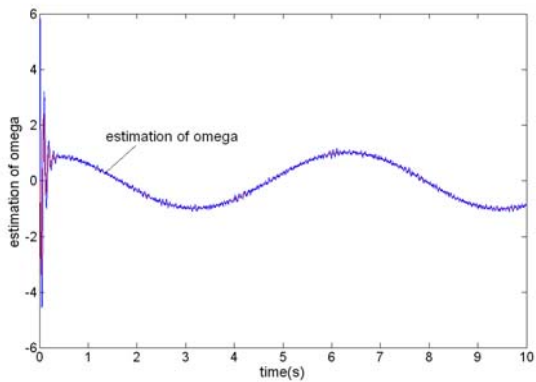

Fig. 4-5 Estimation of $\omega$ by integral-chain differentiator

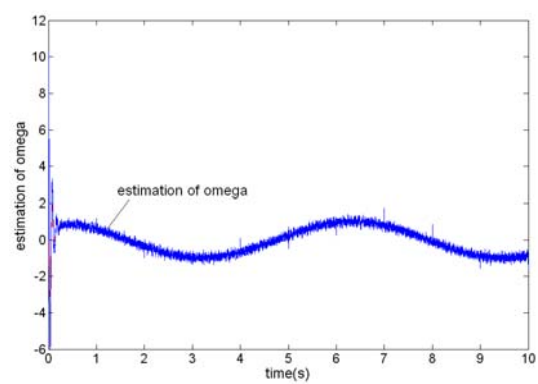

Fig. 5-5 Estimation of $\omega$ by usual high-gain differentiators

# VII. Conclusion

This paper presents the equivalence between integral-chain differentiator and usual high-gain differentiator is given under suitable coordinate transformation. The structure of high-order integral chains in integral-chain differentiator can restrain noises thoroughly. The designed nonlinear integral-chain differentiator is the expansion of linear integral-chain differentiator to increase convergent velocity.

**Acknowledgement**

This paper is supported by National Nature Science Foundation of China (60774008).